# Bayesian reconstruction of sparse raster-scanned mid-infrared optoacoustic signals enables fast, label-free chemical microscopy

Constantin Berger[1,2,3], Myeongseop Kim[1,3], Lukas Scheel-Platz[1,2,3,4], Vasilis Ntziachristos[1,3,5], Dominik Jüstel*[1,2,3], and Miguel A. Pleitez*[1,3]

[1] *Institute of Biological and Medical Imaging, Helmholtz Zentrum München, Neuherberg, Germany*
[2] *Institute of Computational Biology, Helmholtz Zentrum München, Neuherberg, Germany*
[3] *Chair of Biological Imaging at the Central Institute for Translational Cancer Research (TranslaTUM), School of Medicine and Health, Technical University of Munich, Munich, Germany*
[4] *Ludwig Maximilian University of Munich, Munich, Germany*
[5] *Munich Institute of Robotics and Machine Intelligence (MIRMI), Technical University of Munich, Munich, Germany*

*Corresponding authors: dominik.juestel@tum.de, miguel.pleitez@tum.de

**Abstract.** Hyperspectral optoacoustic microscopy (OAM) enables obtaining images with label-free biomolecular contrast, offering excellent perspectives as a diagnostic tool to assess freshly excised and unprocessed tissues. However, time-consuming raster-scanning image formation currently limits the translation potential of OAM into the clinical setting—for instance, in intraoperative histopathological assessments—where micrographs of excised tissue need to be taken within a few minutes for fast clinical decision-making. Here, we present a non-data-driven computational framework tailored to enable fast OAM by sparse data acquisition and model-based image reconstruction, termed Bayesian raster-computed optoacoustic microscopy (BayROM). Unlike conventional machine learning, BayROM doesn't require training datasets, but instead, it employs 1) optomechanical system properties to define a forward model and 2) prior knowledge of the imaged samples to facilitate reconstructing images based on the sparsely acquired data. We show that BayROM enables acquiring micrographs ten times faster and with structural similarity (SSIM) indices greater than 0.93 compared to conventional raster scanning microscopy, thus facilitating the clinical translation of OAM for fast, label-free intraoperative histopathology.

## 1 Introduction

Intraoperative histopathological examinations are crucial for precise surgical decision-making, for instance, in tumor margin analysis [1]. The clinically established workflow for histological assessment usually consists of immunohistochemical (IHC) or hematoxylin-eosin (H&E) stains applied to excised tissue samples, followed by an examination via bright field microscopy [2]. Tissue staining uncovers biomolecular features required for histopathological examinations such as tumor margin assessments. However, it can delay or hinder surgical decision-making due to the time-consuming and laborious preparation steps required [3]. Additionally, while current fast histology techniques used in, for instance, intraoperative tumor margin assessment are rapid (ca. 10 min), they often come with the risk of inaccuracies as they lack the necessary molecular specificity of conventional histology examinations [4]. As an alternative to conventional histology, label-free optoacoustic microscopy (OAM) has been proposed as a method for rapid intraoperative histology due to its ability to obtain intrinsic molecular contrast that allows avoiding the tissue preparation steps usually required for exogenous staining [5, 6]. Particularly, optoacoustic hyperspectral imaging—i.e., sequential imaging at multiple excitation wavelengths over a spectral range—is key for tissue classification based on spectral hallmarks. However, although label-free OAM for histological examination saves time in tissue preparation, acquiring optoacoustic micrographs is often more time-consuming than routinary used bright field microscopy as image formation in many OAM systems is achieved by point-by-point raster scanning. Although technologies for wide-field OAM were proposed to bypass the need for time-consuming raster scanning, these methods often suffer from poor spatial resolution and shallow imaging depth [7]. Low imaging speed is one of the major limiting factors of OAM towards its implementation as an intraoperative histopathological assessment procedure, especially for hyperspectral imaging where time cost scales linearly with the number of excitation wavenumbers acquired [3]. Hyperspectral optoacoustic image formation with fast data acquisition—i.e., within a time range of a few minutes—is crucial to allow for faster surgical decision-making than conventional tissue staining methods [8].

To facilitate fast label-free OAM, several hardware-based methods have been developed, for instance optoacoustic micro-tomography [9], which allows the acquisition of numerous wide-field image volumes per second. Nevertheless, optoacoustic micro-tomography leads to complex optical forward problems, making image formation more prone to artifacts compared to OAM with single-spot illumination combined with raster scanning. Further hardware-based solutions include accelerated raster scanning by optical setups involving voice coil stages [10], microelectromechanical systems (MEMS) [11, 12, 6], polygonal mirror scanners [13], and galvanometer scanners [14]. However, heterogeneously distributed illumination and/or transducer de-focusing typically bear disadvantages associated with a limited field of view (FOV) of the imaged tissues compared to single-point illumination and focused transducers. Limited FOV areas require mosaicking techniques to achieve full image sizes, which are prone to generating stitching artifacts [15, 8].

A strategy to enhance imaging speed while maintaining the advantages of single-point raster scanning of OAM systems such as mid-infrared (mid-IR) OAM [16] is the use of sparse raster scanning in combination with computational methods. Sparse raster scanning is used to reduce data acquisition time by acquiring only a fraction of all pixel intensities in a FOV of interest. To reconstruct full images, i.e., compensate for the inherent information loss due to sparse data acquisition, data-driven methods involving machine learning (ML) and deep learning (DL) have been applied [17-20]. ML and especially DL are effective tools in the realm of image processing and enhancement [21-24]. However, in learning-based methods, the model architectures often used operate as black-box models and do not provide the necessary transparency to ensure artifact-free images, particularly for out-of-distribution data, i.e., sample types that were not covered by the training data [25-28]. To mitigate the risk of imaging artifacts for a broad range of sample types, DL methods would require large amounts of training data, which are often unavailable or impractical to collect in OAM. Thus, although learning-based computational methods have shown promising capabilities for image reconstruction from sparse data, they cannot ensure artifact-free images for a broad range of sample types due to the lack of big training data in OAM. An alternative to data-driven methods is compressed sensing, where prior information about the tissue is integrated via sparsity assumptions [29, 30]. However, compressed sensing does not provide parameters for uncertainty quantification and, thus, similarly to data-driven methods, cannot deliver a quality assessment for reconstructed images to ensure artifact-free imaging. Hence, an alternative method for fast raster scanning OAM, consisting of a transparent computational method to provide image reconstructions, including a quality-check readout, would push the boundaries of label-free imaging toward its integration into intraoperative workflows, but it remains currently unavailable.

We hypothesized that by implementing sparse data acquisition in combination with model-based image reconstruction, we could accelerate the imaging speed of OAM over conventional raster scanning. Additionally, by applying model-based image reconstruction, we could circumvent the limitations imposed for learning-based methods, i.e., the need for big data and/or the non-transparency of neural networks. For this purpose, we developed a computational imaging method termed Bayesian Raster-computed Optoacoustic Microscopy (BayROM). BayROM facilitates the reconstruction of images based on sparse raster scanning data, not required to train a model, and thus bypasses the need for large amounts of training data in OAM. Instead of implicitly learning priors via a dataset, BayROM utilizes an explicit prior model resembling knowledge about the imaged specimen. To maintain similar image quality as in full raster scanning, BayROM reconstructs images based on sparse data supported by a parameterizable model, including optomechanical system properties by means of Variational Bayes and thereby achieves a 10-fold increase in imaging speed compared to conventional raster scanning in OAM.

Here, we showcase BayROM in the context of fast label-free biomolecular imaging of freshly excised adipose tissue by mid-IR hyperspectral microscopy, demonstrating its capability to reduce imaging time from the timescale of hours to only a few minutes. Hence, BayROM paves the way for clinical validation of hyperspectral OAM and its translation to intraoperative histopathological examinations.

## 2   Results

**Working principle of BayROM**

**Fig. 1** graphically illustrates the working principle of BayROM as compared to conventional raster scanning in OAM. Raster scanning OAM measures the optoacoustic signal intensity in each pixel location to compose an image. BayROM increases imaging speed in OAM by systematically skipping raster scanning lines during optoacoustic signal acquisition and compensates for the missing information by employing Bayesian image reconstruction. Therein, a prior model constrains the reconstruction of unscanned areas to plausible values based on the scanned areas. Our knowledge of the measurement process (encoded in the forward model) and the measurement noise allows us to calculate to what degree a candidate image is compatible with the observed data. Furthermore, a priori knowledge about the imaged specimen is included in the form of a generative prior model, which statistically describes the structural properties of images in the harmonic/Fourier domain (see **Methods**). Based on the prior model and likelihood, we can express the plausibility of an image given the acquired data using Bayes' theorem. Bayes theorem allows assigning a plausibility score (posterior probability) from 0 (the image is incompatible with the data, prior knowledge, or both) to 1 (only this image is compatible with the prior knowledge and the observed data). Based on the mean and standard deviation of the posterior probability distribution, we can generate image reconstructions, including pixel-wise uncertainty estimates. We approximate the posterior distribution using the Metric Gaussian Variational Inference (MGVI) algorithm [31], which generates a set of posterior samples. We subsequently calculate posterior means and standard deviations based on the posterior samples. The reconstructed image is derived from the posterior mean, while the posterior standard deviation gives a lower bound of the pixel-wise uncertainty in the reconstruction, which can be used to obtain a quality control metric of reconstructed images. More details on the theoretic principle of BayROM are given in the **Methods** section. Uncertainty-based quality control facilitates the evaluation of reconstructions without the need to compare to a usually unavailable ground truth measurement. Therefore, we use the mean relative standard deviation (MRSD) of the posterior distribution as a metric to reflect the reconstruction uncertainty. Using the MRSD, our Bayesian imaging framework allows for quality-controlled image reconstruction from sparse raster scanning to enable fast optoacoustic imaging.

**Imaging characterization and evaluation in synthetic samples**

First, we characterized BayROM's accuracy by comparing BayROM reconstructed micrographs of a synthetic test target (carbon tape, see **Methods** for details) with full raster scanning images, used as ground truth, of the same sample. **Fig. 2a** shows ground truth and corresponding BayROM reconstructed micrographs (2x2 mm$^2$ FOV and 5 µm pixel size) for the test target at 2850 cm$^{-1}$ excitation. While full raster scanning (ground truth) required 23.43 min for acquisition, data acquisition with 92.5% sparsity, i.e., only 7.5% of the amount of data needed for full raster scanning, required only 2.15 min—thus achieving ca. a 10-fold speedup with an excellent reconstruction accuracy as determined by a SSIM of 0.978 between ground truth and BayROM reconstructed micrographs. **Fig. 2a** brings more details on the reconstruction results, including a backprojection image shown to visualize the useable information contained in the sparse data (see **Fig. 2b**). Furthermore, **Fig. 2a** exhibits the posterior distribution represented by the posterior samples, posterior mean, and posterior standard deviation. The pattern of the posterior standard deviation follows the sample structure, which reflects localization uncertainties related to the point spread function (PSF). Such localization uncertainties can mainly be found at sharp intensity edges since the PSF leads to smoothed scanning data that represents such edges and thus causes uncertainties in their reconstruction. In addition to quantitatively assessing the reconstruction accuracy using the SSIM, **Fig. 2c** exhibits a qualitative comparison using overlay and cross-sectional intensity plots with no noticeable deviations, confirming accurate reconstruction. **Fig. 2d** shows that the reconstruction intensities (i.e., the values of the micrograph after reconstruction) cover the full dynamic range of the ground truth and follow an almost linear power-law relationship with the ground truth values—which is important for quantitative imaging where contrast levels in a micrograph indicate optical absorption of the sample at a given wavenumber. Furthermore, **Fig. 2e** visualizes the distribution of the pixel-wise relative standard deviation (RSD), i.e., the ratios between the posterior standard deviation and the posterior mean for all pixels in the image. The overall uncertainty given by the mean RSD (MRSD) amounts to 4.7% and thus indicates high confidence in the algorithm to reconstruct the images.

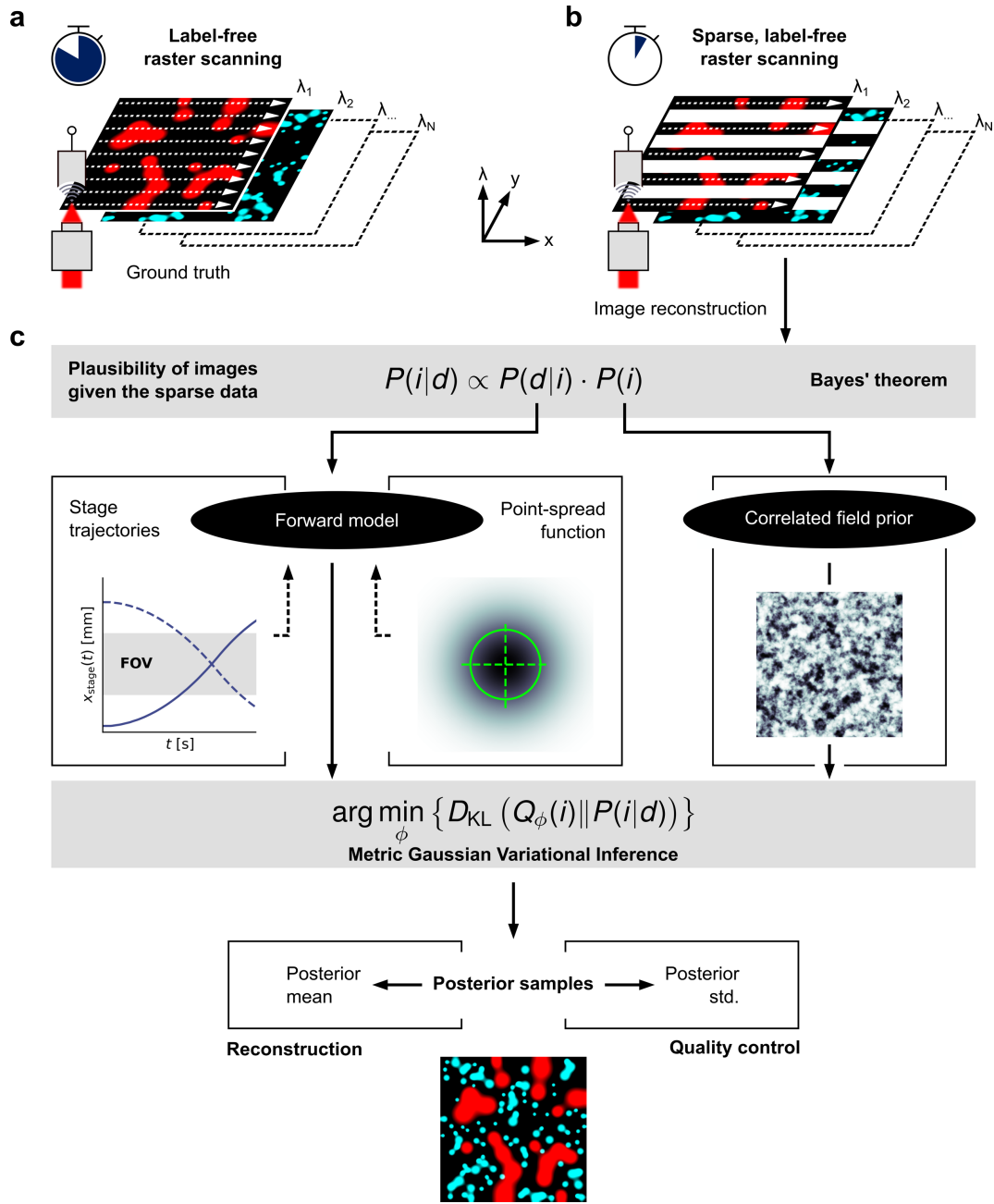

**Fig. 1. Fast label-free imaging using sparse raster scanning and Bayesian image reconstruction. a,** OAM workflow. Raster-scanning OAM enables label-free imaging of unprocessed, freshly excised tissue samples. However, raster scanning is a tedious procedure that prevents fast intraoperative decision-making due to time-consuming data acquisition. **b,** Workflow of Bayesian raster computed optoacoustic microscopy (BayROM). Sparse raster scanning can be applied to decrease the data acquisition time by a factor of 10. Image reconstruction is required to obtain high-resolution micrographs based on the compressed data. **c,** Image reconstruction workflow. The probabilistic reconstruction framework creates plausible images given sparsely acquired data. The plausibility of image candidates can be assessed using the Bayesian posterior probability P(i|d), which is proportional to the product between the likelihood P(d|i) and the prior P(i). The likelihood evaluates the compatibility of image candidates with observed data given by the forward model. The prior model (correlated field prior) compensates for the information loss due to sparse scanning. The mean of the posterior approximated using MGVI serves as the reconstructed image, while the approximated posterior variance provides a metric for quality control.

Next, we studied the effects of varying sparsity parameters on the reconstruction quality to determine a suitable sparsity level with optimal image quality and imaging speedup. The data sparsity is mainly determined by the lateral sparsity parameter, which describes how many raster lines are skipped to increase the data acquisition speed. The data acquisition speed scales linearly with the level of lateral sparsity, i.e., if half of the raster lines are being skipped, data acquisition will be twice as fast. To generate the results shown in **Fig. 2a-e**, 3 out of 4 raster lines were skipped compared to full scanning, i.e., 75% lateral sparsity. With the given sparsity levels, the distances between the raster lines remain smaller than the imaged structures, such that the relevant structural information of the imaged sample is still captured while the imaging speed is increased. The lateral sparsity of 75% leads to an overall data sparsity of 92.5% since, in addition to lateral sparsity, we decreased the averaging level per pixel compared to full raster scanning. High averaging per pixel causes slow stage movements along the imaged raster lines, thus we decreased the averaging level (longitudinal sparsity) so that the stage could operate at maximum speed. To study the effects of changing the number of skipped lines, we analyzed lateral sparsity levels of 50% (every 2nd raster line imaged), 75% (every 4th raster line imaged), and 87.5% (every 8th line skipped), which lead to overall sparsity levels of 85%, 92.5%, and 96.6%. The speedup resulting from the sparse data acquisition depends on the level of sparsity and increases with increasing sparsity levels. At the same time, image quality decreases with increasing sparsity levels. We compared the reconstruction qualities corresponding to sparsity levels of 85%, 92.5%, and 96.6%, leading to acquisition speedup factors of approximately 5, 10, and 20, respectively. **Fig. 2f** visualizes the error distributions of the reconstructions, referring to the pixel-wise absolute errors between reconstructions and ground truth, corresponding to each sparsity level. The analysis of the reconstruction error distribution with respect to the sparsity level shows that the mean reconstruction error remains similar between the compared sparsity levels. However, the maximum reconstruction errors increase significantly with increasing sparsity. Generally, the choice of a sparsity level and, thus, the resulting speedup constitutes a tradeoff in image quality and imaging speed that strongly depends on the application and thus needs to be seen as a parameter of BayROM and chosen accordingly. The results displayed in **Fig. 2** demonstrate the capabilities of BayROM to accurately reconstruct sparse images under consideration of varying sparsity parameters and inherent imaging speedups. Moving towards clinical applications, BayROM can be applied to biological tissues to further assess its capabilities with respect to reconstruction quality and imaging time.

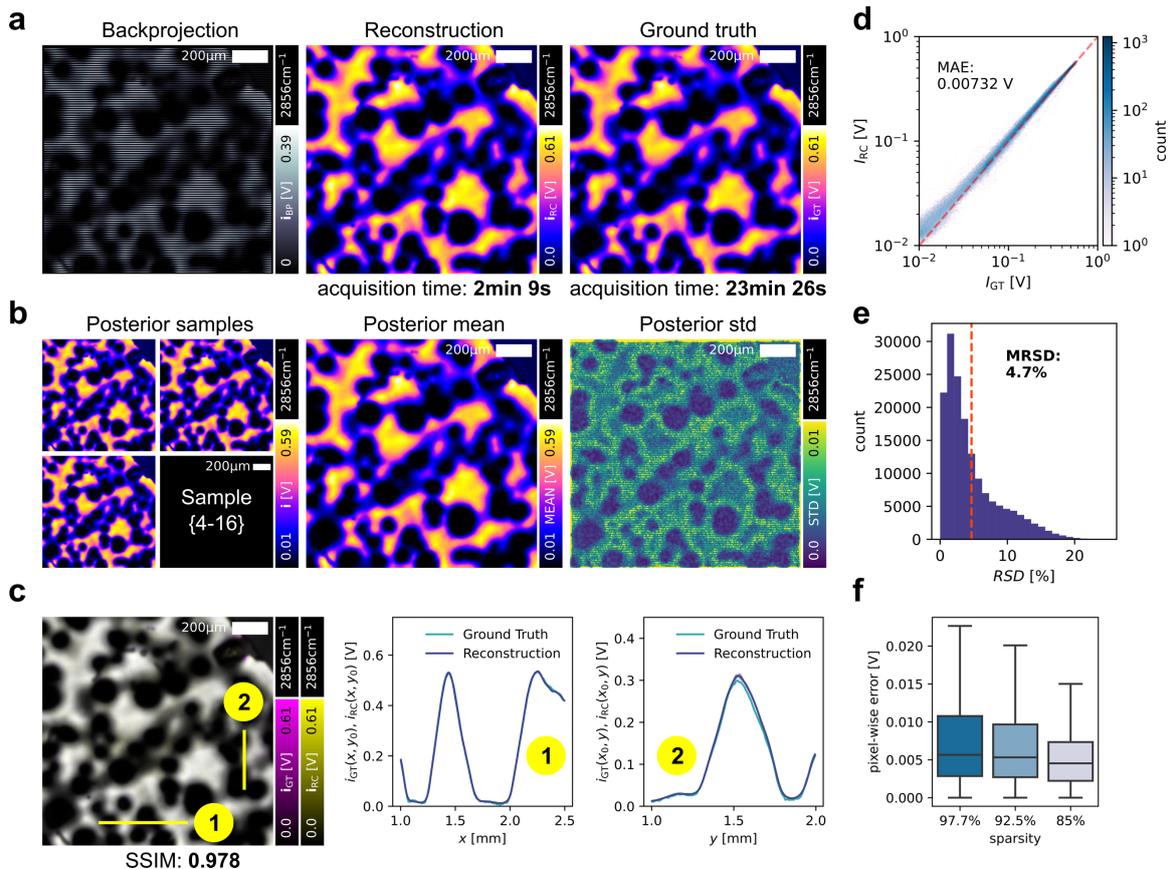

**Fig. 2. Reconstruction analysis and characterization on a test phantom. a,** Visual comparison of back-projected data, reconstruction, and ground truth. The backprojection shows a horizontal line pattern, which originates from the lateral data sparsity. The SSIM between reconstruction and ground truth is 0.978. **b,**

Visualization of the posterior distribution. The posterior mean and variance are obtained based on a pixel-wise assessment of the posterior samples. The posterior standard deviation shows a horizontal line pattern, which reflects higher uncertainties in skipped lines due to lateral sparsity during data acquisition and increased uncertainty on the boundaries of tissue structures due to PSF-related localization uncertainty. **c,** Overlay of reconstruction and ground truth. The overlay, as well as the cross-sectional line plots in both the x and y directions, confirm an almost perfect reconstruction of the ground truth. **d**, Intensity comparison between reconstruction and ground truth. The distribution of intensities in the ground truth image versus the reconstructed image is well aligned on the centerline, meaning that there is negligibly low bias in the intensity profile of the reconstruction. **e**, Histogram of the pixel-wise relative standard deviation (RSD). The mean relative standard deviation (MRSD) suggests an average reconstruction uncertainty of 4.7%. **f**, Sparsity analysis. A change in the level of lateral sparsity results in a similar mean of the pixel-wise error between reconstruction and ground truth. However, the error distribution gets expanded, as indicated by the bars. An overall sparsity level of 92.5% was used to generate the results shown in **a-e**.

**Fast hyperspectral imaging of unprocessed white adipose tissue (WAT)**

To demonstrate the capabilities of BayROM for fast imaging of biological specimens, we assessed the imaging speed as well as the reconstruction accuracy of images obtained for white adipose tissue (WAT) excised from mice at multiple excitation wavenumbers (i.e., at multiple contrast channels). Specifically, we imaged fixed epididymal WAT taken from Friend leukemia virus B mice for a FOV of 1x1 mm$^2$ and a pixel size of 2.5 µm at selected wavenumbers, including 2856 and 1550 cm$^{-1}$, which provide mainly contrast for lipids and proteins, respectively. **Fig. 3** shows more details about the imaging performance, including an overlay of the reconstructions for each channel compared with the corresponding ground truths. **Fig. 3c** shows cross-sectional intensity plots of the ground truth and reconstructed image in both the x- and y- directions, demonstrating accurate reconstruction. **Fig. 3d** shows the ground truth and reconstruction intensity values of all excitation wavenumbers in a scatterplot, indicating no noticeable biases, thus suggesting linear behavior between ground truth and reconstructed pixel intensities in the entire value range. For an overall data sparsity of 92.5%, i.e., 3 out of 4 raster lines skipped, BayROM achieves approximately 10 times faster data acquisition while maintaining high image quality expressed by a SSIM of 0.950 by comparison with the ground truth. Similar to the characterization based on the carbon sample, we analyzed different sparsity levels, i.e., 85%, 92.5%, and 96.6%, to assess their corresponding performances. **Fig. 3e-f** shows reconstructed hyperspectral images of gonadal white adipose tissue imaged in a FOV in a 1x1 mm$^2$ with a pixel size of 2.5 µm. While all sparsity levels lead to SSIMs above 0.8, the sparsity of 96.6% shows that the structures of the adipocytes are becoming less recognizable in some parts of the image (SSIM: 0.805). While the 85% sparsity results in the highest SSIM value of 0.950, only a speedup factor of approximately 5 was achieved. Thus, imaging at a sparsity level of 92.5% offers, with a speedup factor of approximately 10 and a SSIM of 0.934, the best compromise between imaging speed and minimal deviations from ground truth. In this way, BayROM imaging enables higher information throughputs than full raster scanning, i.e., imaging more channels over time and thus facilitating fast hyperspectral imaging, i.e., the sequential image acquisition at different excitation wavenumbers, which is specifically relevant to assess the biomolecular composition of imaged specimens.

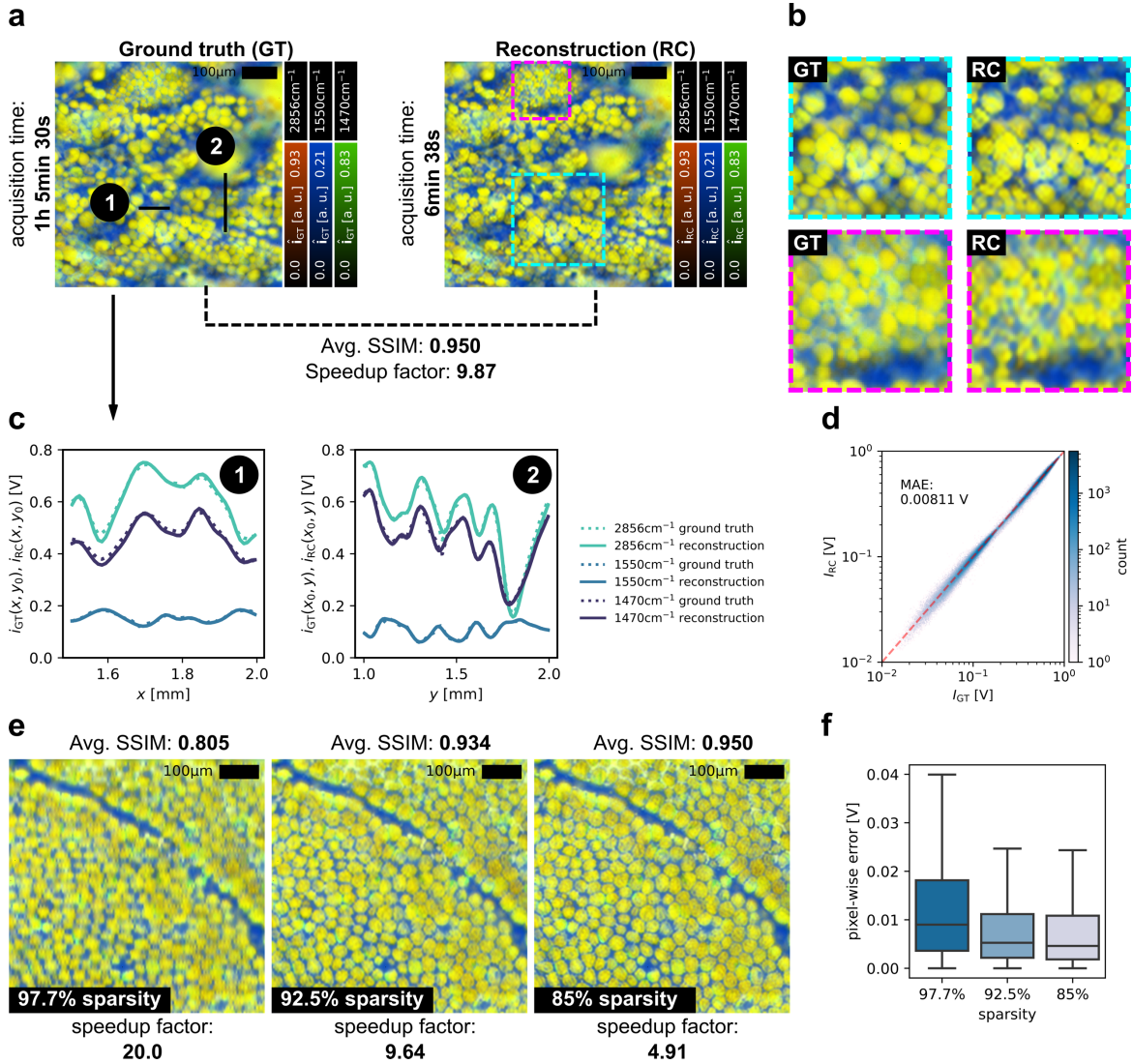

**Fig. 3. BayROM applied to adipose tissue imaging. a,** Ground truth hyperspectral image of white adipose tissue obtained using full raster scanning and reconstructed hyperspectral image acquired using sparse raster scanning. The BayROM image was acquired with approximately 10 times faster data acquisition speed compared to the ground truth reference using raster scanning. Despite the speed increase of one order of magnitude, the average structural similarity index (SSIM) between reconstructed and ground truth images is 0.950, confirming the high fidelity of the fast scanning and reconstruction method. **b**, Visual comparison between ground truth and BayROM image regarding two selected areas in the image. **c**, Qualitative comparison of reconstruction and ground truth. The intensities for each excitation wavenumber in the cross-sectional intensity plots in both the x- and y-direction show that the BayROM image accurately follows the ground truth intensities. **d**, Quantitative comparison of reconstruction and ground truth. The error plot shows a mean absolute error of 0.00811V. Moreover, the error plot confirms that there is no bias in the reconstructed intensities as the scatter plot follows the diagonal line. **e**, Comparison of sparsity levels. A sparsity level of 92.5% was chosen as it enables high reconstruction quality and approximately 10 times faster data acquisition. **f**, Comparison of pixel-wise errors for sparsity levels shown in e.

Finally, having achieved the ability to acquire images ten times faster than with conventional raster scanning, we moved on to apply BayROM for hyperspectral imaging of freshly excised tissues. This was demonstrated by sequential raster scanning a WAT sample for a FOV of 1x1 mm$^2$ and 5 µm pixel size at 80 different excitation wavenumbers in two spectral ranges covering lipid (2800 – 2925 cm$^{-1}$) and protein contrast (1500 – 1700 cm$^{-1}$). Based on an overall sparsity level of 92.5%, we reconstructed each channel to compose an image stack. In **Fig. 4a-b,** we visualized the ground truth image of WAT acquired by full faster scanning based on two channels (2856 and 1550 cm$^{-1}$) as well as the corresponding 80-channel BayROM image stack, denoted as hypercube. The hypercube contains the complete spatial and spectral information of the sample, i.e., each pixel on the hypercube comprises the mid-IR spectrum for the corresponding location in the sample. The spectral stability of hypercubes generated using BayROM—i.e., the accuracy of reconstructed spectra obtained from spatially sparse data acquisition—was assessed by comparing the spectra obtained with the reconstructed hypercube at selected pixels with the ground truth spectra acquired from the same locations. We visualized the spectra from the BayROM hypercube in comparison with the ground truth spectra in **Fig. 4d-g**, confirming that BayROM achieves accurate reconstructions of spectra, quantified by the mean relative error (MRE), which amounts to 4.4% for the adipocyte spectra and 1.7% for the ECM spectra. To further analyze the imaged tissue, we used the ground truth spectra taken from the locations marked in **Fig. 4a** representing adipocytes and extracellular matrix (ECM) and performed linear spectral unmixing to generate a spatial mapping of the resulting mixing coefficients visualized in **Fig. 4c**. The spatial mapping of unmixing coefficients can be used to assess the biological composition of the imaged sample, which according to the unmixing consists of 66.4% adipocytes and 33.6% ECM. In summary, we confirmed the hyperspectral imaging capability of BayROM, which is crucial for analytic downstream tasks, i.e., methods for data analysis applied after imaging, such as biological assessments based on spectral unmixing.

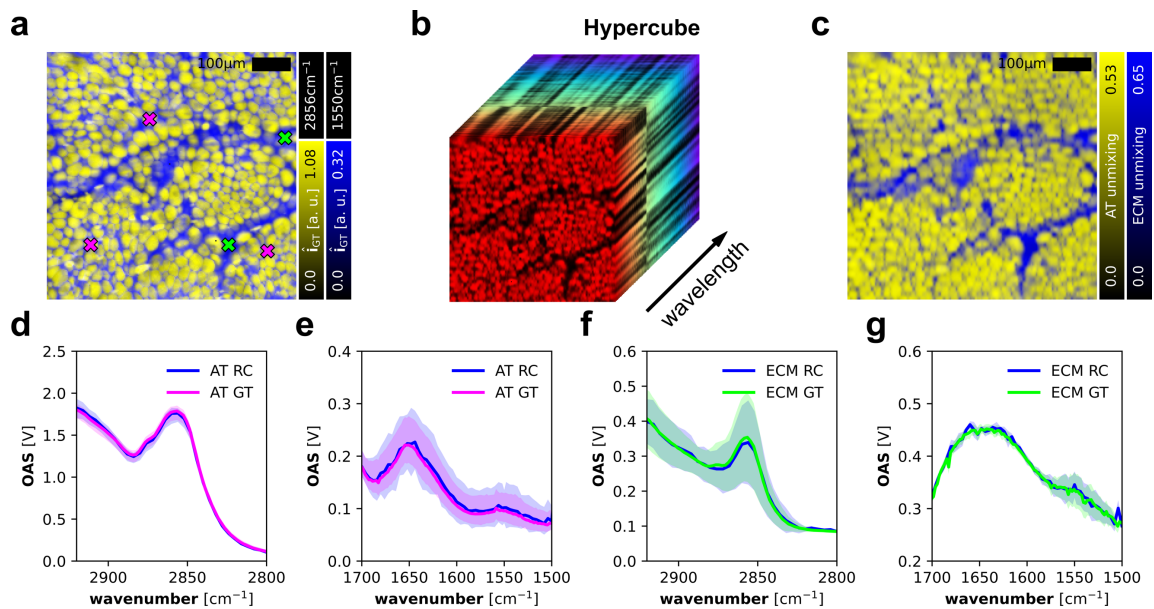

**Fig. 4. Hyperspectral adipose tissue imaging. a,** White adipose tissue imaged with full raster scanning. Crosses mark selected positions representing adipocytes (AT, pink) and extracellular matrix (ECM, green). **b**, Hyperspectral BayROM image. The hypercube consisting of 80 wavenumbers was captured at the same FOV as in **a**. **c**, Pixel-wise linear unmixing of AT and ECM spectra. According to the pixel-wise unmixing coefficients, the imaged tissue consists of 66.4% adipocyte content and 33.6% ECM. **d**, Comparison of AT spectra in the lipid region. **e**, Comparison of AT spectra in the amid region. **f**, Comparison of ECM spectra in the lipid region. **g**, Comparison of ECM spectra in the amid region. The reconstructed spectra were obtained from the pixel intensities in the BayROM hypercube corresponding to the locations where the ground truth spectra were acquired. The shaded areas in **d-g** represent the ranges between the maximum and minimum spectral intensities.

## 3 Discussion

We showed that sparse data acquisition in OAM combined with Bayesian image reconstruction using BayROM leads to approximately ten times faster imaging speed compared to full raster scanning while preserving image quality. Specifically, we demonstrated on synthetic samples and adipose tissue from mice that BayROM generates accurate reconstructions when compared to ground truth images acquired using full raster scanning (SSIM greater than 0.93). Raster scanning OAM is considered a low-throughput imaging technique due to its slow imaging speed. Slow imaging speed in OAM is a key limitation preventing its translation into clinical workflows despite its proven label-free molecular imaging capabilities. BayROM could help to overcome this limitation by accelerating data acquisition while avoiding artifacts related to hardware-based solutions for fast raster scanning and avoiding the need for large data sets required by learning-based approaches that involve neural networks.

Model-based image reconstruction by BayROM is a non-data-driven approach that compensates for the sparse acquisition of data by using a generative prior model. Unlike learning-based approaches, BayROM does not require domain-specific training datasets to reliably compensate for missing information without the risk of generating artifacts. This main advantage of BayROM makes the tedious and often infeasible collection of large training data sets in OAM unnecessary. Additionally, shifting from "black box" DL-based reconstruction approaches to Bayesian image reconstruction offers more transparency by incorporating quality control parameters, i.e., the reconstruction uncertainty, for assessing the uncertainty associated with reconstruction results. However, image reconstruction with BayROM requires more computational effort/time than DL-based solutions. A high computational effort is required because model-based image reconstruction by BayROM is achieved using MGVI, a variational inference algorithm that solves a high-dimensional optimization problem. Contrary to DL-based solutions, where image reconstruction could be achieved by single forward propagation processes of neural networks, variational inference is solved in iterative loops. Nonetheless, fast image reconstruction using BayROM could be achieved by 1) increasing the computational power used for image reconstruction, 2) optimizing hardware by using graphics processing units (GPUs) for parallelized implementations of computational operations, etc., or 3) approximate reconstruction, meaning that the posterior distribution is approximated with fewer iterations of the MGVI algorithm, making reconstructions faster than full reconstruction. To showcase the effects of approximate reconstruction, we analyzed the reconstruction accuracy for approximated versus full image reconstruction (**supplementary Fig. 1**). To do this, we used 3 instead of 5 iterations of MGVI combined with 8 instead of 16 samples drawn from the posterior distribution to reconstruct the images presented in **Fig. 3**. Approximate reconstruction was carried out in only 2 min 38 s compared to 10 min 6s needed for full reconstruction. The result shows that even with approximate reconstruction, similar reconstruction residuals can be achieved compared to standard reconstruction while saving approximately 75% of the computational time. However, the maximum uncertainty in the approximate reconstruction is approximately 2-fold compared to the standard reconstruction method, which shows that the approximate solution provides less confidence about the resulting image than the full reconstruction and thus needs to be applied with the awareness that the imaging is potentially less accurate. When combining solutions 1, 2, and 3, the computational time required for model-based reconstruction using BayROM can be substantially lowered to obtain images within a few seconds instead of several minutes.

BayROM provides a parameterizable imaging framework that can be tuned to meet the specific needs of the application. The adjustable parameters include the sparsity setting, i.e., the number of skipped raster lines, as well as the pixel-wise averaging level, and define the imaging speed but also parameters referring to the reconstruction, such as the latent variables of the prior model (see **Methods**). Since the prior model accounts for the missing data points in skipped raster lines, i.e., assists in filling the spatial gaps with structural features that are not or only indirectly contained in the acquired data, the priors can be tuned to represent characteristic sample properties. Altogether, sparsity settings and prior model configuration need to be directly adjusted based on the clinical application, which determines the requirements for both imaging speed and quality. However, although BayROM allows parameterizing the prior model according to the imaged sample types, the prior model might not cover every possible specimen. Samples with sharp transitions orthogonal to skipped raster lines or structures contained in the image that are substantially smaller than the total width of consecutively skipped raster lines are especially challenging and tend to generate artifacts. However, this is generally the case for sparse scanning methods that skip entire raster lines because data needs to be generated when information is entirely missing. To avoid reconstruction artifacts, it is important to choose sparsity parameters such that the acquired data provide enough information about the structures of interest to enable their successful reconstruction. Hence, a thorough choice of parameters has to be made for sparsity levels and latent variables of the prior model in order to match the requirements of the clinical application and potential computational downstream analysis.

The overall benefit of our work is faster OAM imaging based on sparse data acquisition compared to conventional raster-scanning OAM imaging. BayROM provides a parameterizable imaging framework that does not require the collection of training data, unlike related data-driven methods for image reconstruction based on sparse data. Furthermore, BayROM does not suffer from stitching artifacts in larger FOVs, such as common hardware-based solutions for fast scanning. A combination of optomechanical speedup methods with BayROM could potentially enable synergy between the hardware-based and software-based methods to further increase the data acquisition speed towards high-speed and high-throughput OAM applications. On the other hand, the overall limitation of BayROM is the high computational effort required for image reconstruction compared to DL methods. The next step is a hardware-optimized and parallelized implementation to be operated using more computational power in order to assess the imaging performance for clinical applications. A promising clinical application could be tumor margin segmentation based on fast hyperspectral imaging, which would be an important step toward the clinical implementation of BayROM with optimized computation.

In summary, we showed that BayROM, a sparse imaging method combined with a probabilistic image reconstruction algorithm, enables 10 times faster data acquisition and accurate reconstructions, which we showed with adipose tissue imaging. The impact of faster data acquisition is that increased imaging speed of raster scanning OAM by a factor of 10 could enable its integration into surgical workflows, such as label-free intraoperative tumor margin assessment. This is important because if OAM is fast enough, it has the potential to enable faster and more efficient intra-operative decision-making compared to non-label-free gold standard procedures, which require staining or processing of excised tissues. Fast intraoperative decision-making based on label-free molecular imaging and supported by BayROM has the potential to enable more accurate histopathological assessment of excised tissues.

## 4   Methods

**Molecular contrast formation and optomechanical setup**

The ability of biological molecules to convert light to sound via the optoacoustic effect strongly depends on the arrangement and interaction of their constituent atoms and atomic bonds, as well as the wavelength of the light interacting with them. For this reason, biological molecules have characteristic optoacoustic spectral fingerprints, which OAM exploits to generate label-free molecular contrast. The types of molecules present in a sample, their spatial distribution, and local concentrations all shape a sample's ability to generate optoacoustic signals. OAM measures this position- and excitation wavelength-dependent optoacoustic signal generation ability, henceforth called the optoacoustic signal strength field $OASS(x, \lambda)$. Molecular prevalences in tissues can be deduced by identifying molecular fingerprints in the optoacoustic signal strength field.

Mid-infrared optoacoustic microscopy (MiROM) [16], the OAM technique employed in this work, uses a broadly tunable quantum cascade laser (MIRcat, Daylight Solutions) to probe the optoacoustic signal strength field of a sample with mid-IR radiation in the wavenumber range of 2,941-909 cm$^{-1}$ (3.4-11 µm). The sample is placed on a mid-IR-transparent Zink sulfide window (Crystal) and illuminated from below. The laser beam is focused on a plane located in the imaged sample using a 0.5 NA reflective objective (36x, Newport Corporation) to confine the optical excitation to a small tissue volume. To sense the generated optoacoustic signal, a focused ultrasound transducer (Imasonic) with a central frequency of 20 MHz is placed above the tissue and co-aligned to the focal spot of the mid-IR laser, capturing the optoacoustic signals through a coupling medium (deionized water). The optoacoustic signals are acquired using a data acquisition card (Gage Applied) after being amplified by a 63 dB low-noise amplifier (MITEQ) and processed by a 50 MHz low pass filter (Mini-circuits). The mid-IR laser has a repetition rate of 100 kHz and a pulse duration of 20 ns. Three exemplary wavelengths were used to assess the molecular response in the carbon sample and the adipose tissue samples: 2856 cm$^{-1}$ causing symmetric stretching of CH2 functional groups, 1550 cm$^{-1}$ causing N-H bending/C-N stretching, and 1470 cm$^{-1}$ causing CH2/CH3 bending.

Raster scanning OAM maps the optoacoustic signal strength field by spatially raster scanning the sample for each selected excitation wavelength. In each pixel location (indexed by the pixel coordinates $k \in \{1, ..., n\}$ and $l \in \{1, ..., m\}$), the optoacoustic signal strength field is probed with multiple laser pulses (indexed by the pulse number $j \in \{1, ..., 50\}$), yielding a set of optoacoustic transient signals $OAT_{k,l}^j(t)$. These transient signals are subsequently averaged with respect to the laser pulses to increase the SNR, resulting in the averaged optoacoustic transient signal $OAT_{k,l}(t)$. To form a MiROM micrograph

$$\mathbf{i}^* = \begin{pmatrix} i_{1,2}^* & \cdots & i_{1,m}^* \\ \vdots & \ddots & \vdots \\ i_{n,1}^* & \cdots & i_{n,m}^* \end{pmatrix}$$

with $n \cdot m$ pixels (henceforth called „image"), the peak-to-peak amplitudes of the averaged optoacoustic transient signals $OAT_{k,l}(t)$ are extracted as the image pixel values

$$i^*_{k,l} = \max_t \left( OAT_{k,l}(t) \right) - \min_t \left( OAT_{k,l}(t) \right), \quad \forall k \in \{1, \ldots, n\}, l \in \{1, \ldots, m\}.$$

We speed up data acquisition by skipping raster scanning lines $l$ (lateral sparsity) and reducing the number of optoacoustic transient signals acquired in each measurement location from 50 to 15 (longitudinal sparsity). This acquisition approach yields a sparse dataset

$$d_{k,v} = \max_t \left( OAT'_{k,v}(t) \right) - \min_t \left( OAT'_{k,v}(t) \right), \quad \forall k \in \{1, \ldots, n\}, v \in \{1, w+1, 2w+1, \ldots\},$$

where $OAT'_{k,v}(t)$ denotes the averaged optoacoustic transient signals formed from 15 laser pulses and $w$ denotes the raster scanning line acquisition stride. When measuring every 4th raster scan line ($w = 4$), the amount of data acquired reduces by 92.5% with respect to a full scan.

The sparse measurements permit a significant speed-up of the data acquisition process but are realized at the cost of information loss with respect to the baseline. The increased stride between raster scanning lines creates areas in the micrograph where the optoacoustic signal strength field is not directly probed, and the reduction in repeated transient measurements leads to a reduced SNR in the data **d**. Consequently, reconstructing dense micrographs **i** based on the sparse data **d** is an ill-posedness inverse problem (i.e., does not have a unique solution), which we address in the framework of Bayesian image reconstruction.

**Exploitation of structural regularities in imaged samples**

Many samples of interest for OAM have characteristic structural regularities, such as the grain structure of carbon tape or the arrangement, sizes, and shapes of adipocytes in white adipose tissue. The core concept of BayROM is that OAM images can be reconstructed faithfully from sparse data if the imaged samples exhibit structural regularities, which can be used to deduce the image values in areas without measurements from measurements in other areas. For example, knowledge of the typical shape and size of adipocytes in white adipose tissues allows predicting the full contour of a partially imaged adipocyte.

In general, i.e., for arbitrary samples, such structural regularities are not known apriori. Nevertheless, to reconstruct a broad range of samples, BayROM estimates the structural regularities of the imaged sample during the image reconstruction process based on the available data **d** and forms the image using the estimated regularities. For this purpose, we model the image as a (non-linearly transformed) Gaussian random field with approximately Matérn covariance structure. The covariance structure encodes the structural regularities of the image and is inferred in the image reconstruction process. Although exploiting the structural regularities of the sample constrains the solution space, the imaging problem remains ill-posed.

**Bayesian image reconstruction**

Given there is no unique solution to the image reconstruction problem, we pursue a probabilistic image reconstruction approach. We recover the discrete posterior probability distribution of images $p(\mathbf{i}|\mathbf{d})$, which expresses how likely it is that a full OAM scan would produce the image **i** is given that a sparse scan has produced the data **d**. Based on the posterior distribution of images, we can estimate the expected image for a full OAM scan (via the posterior mean) and the uncertainty of this estimate (via the posterior variance).

Following Bayes' theorem the posterior probability of an image **i** is proportional to the product of the so-called likelihood $p(\mathbf{d}|\mathbf{i})$ and prior probability distribution $p(\mathbf{i})$

$$p(\mathbf{i}|\mathbf{d}) \propto p(\mathbf{d}|\mathbf{i}) \cdot p(\mathbf{i}).$$

The likelihood encodes how probable it is to observe the data **d** for a sample that a full scan would produce the image **i,** given our knowledge about the measurement process. The prior probability distribution encodes our a-priori knowledge about studied samples and their associated OAM images, such as the fact that the images are strictly positive-valued. We also encode our expectation that the images will have relevant structural regularities in the prior.

Computing posterior statistics given $p(\mathbf{i}|\mathbf{d})$ is challenging because of the high dimensionality of the images. We utilize variational inference to approximate the posterior distribution $p(\mathbf{i}|\mathbf{d})$ with a parametric distribution $Q_\phi(\mathbf{i})$, what is constructed so that efficient computation of its distribution statistics is possible. The posterior mean and standard deviation can then be estimated using $Q_\phi(\mathbf{i})$. Variational inference approximates the true posterior distribution by minimizing the Kullback-Leibler (KL) divergence

$$D_{\text{KL}}(Q_\phi||p) = \sum_{\mathbf{i}} Q_\phi(\mathbf{i}) \log\left(\frac{Q_\phi(\mathbf{i})}{p(\mathbf{i}|\mathbf{d})}\right)$$

between the posterior and the variational distribution $Q_\phi(\mathbf{i})$. We use an algorithm optimized for high-dimensional spaces, Metric Gaussian Variation Inference (MGVI) [31], which represents the variational distribution $Q_\phi(\mathbf{i})$ as a vector of latent samples $\phi = (\psi_1, \psi_2, \ldots, \psi_{n_{\text{MGVI}}})$. The image reconstruction process was implemented in the probabilistic inference framework „NIFTy" [32, 33] and was executed on an M2 MacBook Pro (Apple).

**System forward model and likelihood definition**

We model the measurement process of sparse OAM imaging with the combination of a deterministic system response $\mathbf{R}$ and additive stochastic noise $\mathbf{n}$

$$\mathbf{d} = \mathbf{R}(\mathbf{i}) + \mathbf{n}.$$

The deterministic system response includes, on the one hand, the effect of the PSF of the system and, on the other hand, the choice of measurement locations implied by the stage trajectories. The system PSF was characterized based on an image taken from a 1 µm (sub-resolution) polystyrene sphere [16]. Numerically, the operator $\mathbf{R}$ is implemented as sequentially applying a PSF convolution operator $\mathbf{R}_{\text{PSF}}$ (which convolves the image $\mathbf{i}$ with a 2D Gaussian kernel with a standard deviation of 5µm in both spatial dimensions) and $\mathbf{R}_{\text{stages}}$, which models the lateral distribution of measurement locations and selectively copies values of the PSF-convolved image $(\mathbf{R}_{\text{PSF}}\mathbf{i})_{k,l}$ to entries of the simulated data vector $\mathbf{d}$. The additive noise $\mathbf{n}$ was characterized based on the dark noise of the system, which is measured by acquiring an image while blocking the QCL beam and assessing the distribution of the pixel intensities. The dark noise was found to be independent and identically distributed Gaussian noise with a standard deviation of $\sigma = 9.318\text{e}^{-7}\text{V}$. In summary, the sparse OAM measurement is modeled as

$$\mathbf{d} = \mathbf{R}\mathbf{i} + \mathbf{n} = (\mathbf{R}_{\text{stages}} \circ \mathbf{R}_{\text{PSF}})\mathbf{i} + \mathbf{n}.$$

Accordingly, the likelihood $p(\mathbf{d}|\mathbf{i})$ is given by

$$p(\mathbf{d}|\mathbf{i}) = G(\mathbf{d} - R(\mathbf{i})|\boldsymbol{\mu} = \mathbf{0}, \mathbf{N} = \sigma^2\mathbf{I}),$$

where $G$ is a Gaussian distribution with mean $\boldsymbol{\mu} = \mathbf{0}$ and covariance $\mathbf{N} = \sigma^2\mathbf{I}$.

**Image prior implementation**

Following [31], we implement the image prior as a hierarchical generative model $\mathbf{i} = f(\boldsymbol{\psi})$, which deterministically transforms latent parameters $\boldsymbol{\psi}$ into images $\mathbf{i}$. As stated in the section **Exploitation of structural regularities in imaged samples**, the images are modeled as a Gaussian random field GF with approximately Matérn covariance structure, which is non-linearly transformed to model the strict positivity of OAM images

$$f(\boldsymbol{\psi}) = r \cdot \text{sigmoid}\big(\text{GF}(\boldsymbol{\psi}^{\text{GF}}) + s(\psi^s)\big).$$

Here, $r$ is a constant scaling factor setting the maximal obtainable image value. The sigmoid function ensures positivity of the generated images, while $s$ is an additive offset that controls the average pixel value. The optimal additive offset for a given data vector $\mathbf{d}$ is inferred during reconstruction and a-priori follows a Gaussian distribution with mean $\mu_s = 0.5$ and standard deviation $\sigma_s = 0.25$.

The Gaussian random field $\text{GF}(\boldsymbol{\psi}^{\text{GF}})$ is meant to capture and exploit the correlation structure of the image as described above. It is constructed via the Hartley amplitude field $\mathbf{A}(\boldsymbol{\psi}^{\text{GF}})$, which is a harmonic representation similar to Fourier mode amplitudes based on the Hartley transform (HT)

$$\text{GF}(\boldsymbol{\psi}^{\text{GF}}) = \text{HT}\Big(\mathbf{A}(\boldsymbol{\psi}^{\text{GF}})\Big).$$

We model **A** as being separable into a length-scale dependent Hartley spectrum function $E(|\mathbf{k}|)$ and a Gaussian random field $\boldsymbol{\psi}^{\xi}$. Here, $\mathbf{k} = (k, l)$ is the wavevector indexing the Hartley modes. The Hartley spectrum function captures the smoothed spectral power distribution in **A**, while the Gaussian random field $\boldsymbol{\psi}^{\xi}$ captures deviations of the Hartley amplitudes **A** from $E$. Correspondingly, the Hartley amplitudes $A_{k,l}$ is defined as

$$A_{k,l} = E_{k,l}(\psi^a, \psi^b, \psi^c) \cdot \psi^{\xi}_{k,l}.$$

$E(|\mathbf{k}|)$ is constructed as a Matérn spectrum with learnable parameters $a(\psi^a)$, $b(\psi^b)$, and $c(\psi^c)$

$$E_{k,l} = a\left(1 + \left(\frac{\sqrt{k^2 + l^2}}{b}\right)^2\right)^{\frac{c}{4}}.$$

This formulation of GF allows us to incorporate the a-priori expectation of Matérn covariance for the image while permitting the reconstruction to produce images with covariance structures deviating from the Matérn form if the data suggests it. The priors for the Hartley spectrum function parameters $a$, $b$, and $c$ are chosen weakly informative to minimize the risk of biasing the reconstruction.

In summary, the hierarchical image model $f(\boldsymbol{\Phi})$ non-linearly transforms latent parameter vectors $\boldsymbol{\psi} = (\boldsymbol{\psi}^{\xi}, \psi^a, \psi^b, \psi^c, \psi^s)$ into images **i**, which a-priori exhibit Matérn covariance. Inserting $\mathbf{i} = f(\boldsymbol{\psi})$ into the measurement equation

$$\mathbf{d} = \mathbf{R}\mathbf{i}(\boldsymbol{\psi}) + \mathbf{n} = \left(\mathbf{R}_{\text{stages}} \circ \mathbf{R}_{\text{PSF}} \circ f\right)\boldsymbol{\psi} + \mathbf{n},$$

the image reconstruction problem can be reformulated as reconstructing the posterior distribution of the latent parameters $p(\boldsymbol{\psi}|\mathbf{d})$, which has practical benefits as outlined in [34].

**Postprocessing**

The outcome of the MGVI, i.e., a set of posterior samples $\boldsymbol{\phi} = (\boldsymbol{\psi}_1, \boldsymbol{\psi}_2, ..., \boldsymbol{\psi}_{n_{\text{MGVI}}})$, is used to compute the posterior mean image $\mathbf{i}_{\text{RC}}$

$$\mathbf{i}_{\text{RC}} = \frac{1}{n_{\text{MGVI}}} \sum_{i=1}^{n_{\text{MGVI}}} f(\boldsymbol{\psi}_i).$$

$\mathbf{i}_{\text{RC}}$ serves as reconstructed images after applying additional post-processing steps. Post-processing includes cropping the edges of the image, which were added to the sparse micrographs in the numerical convolution with the PSF. Furthermore, if needed, the reconstructed images are interpolated using linear interpolation to allow pixel-wise comparison of the reconstruction based on sparse data with the ground truth images. For visualization, contrast-limited adaptive histogram equalization (CLAHE) was applied to white adipose tissue micrographs to emphasize image contrast. Images **i** that were processed with CLAHE are indicated as **î**. The reported computational times include the full image generation processes, i.e. reconstruction and postprocessing. However, the indicated imaging times do not include the computational time.

**Hyperspectral linear unmixing**

For a demonstration of spectral unmixing based on the hypercube acquired using BayROM, linear unmixing was performed pixel-wise. Therefore, the spectra obtained using single-point optoacoustic spectroscopy were grouped into AT and ECM and averaged. The model for linear unmixing assumes that each pixel's spectrum is composed of a linear combination of both the AT and the ECM average spectrum. Therefore, the mixing coefficients were restricted to non-zero and determined using pixel-wise least squares minimization. The result provides two coefficients for each pixel, which were mapped to the image using an overlay to visualize the unmixing result.

**Evaluation metrics**

The quantitative evaluation of BayROM is two-fold. Since a tradeoff consists of the amount of data acquired and, thus, the acquisition speed and image quality, the two metrics used are, on the one hand, the speedup factor denoting the ratio between data acquisition time of the non-sparse and sparse images and on the other hand the structural similarity (SSIM) between the reconstructed images and the ground truths. The SSIM is defined as

$$SSIM = \frac{(2\mu_{RC}\mu_{GT} + c_1)(2\sigma_{RC,GT} + c_2)}{(\mu_{RC}^2 + \mu_{GT}^2 + c1)(\sigma_{RC}^2 + \sigma_{GT}^2 + c_2)},$$

where $\mu_{RC}$ and $\mu_{GT}$ are the mean values of the reconstructed ground truth images, $\sigma_{RC}$ and $\sigma_{GT}$ are the variances of the reconstructed ground truth images and $\sigma_{RC,GT}$ is the co-variance between reconstruction and ground truth. The variables $c_1 = 0.0001$ and $c_2 = 0.0009$ stabilize the division by small denominators.

**Uncertainty quantification**

To quantify the overall uncertainty in reconstructed images, the mean relative standard deviation (MRSD) was used. The MRSD is calculated based on the estimated posterior mean $i_{k,l}^{RC}$ and standard deviation $u_{k,l}^{RC}$ for each pixel

$$MRSD = \frac{1}{m \cdot n} \sum_{k=1}^{n} \sum_{l=1}^{m} \frac{u_{k,l}^{RC}}{i_{k,l}^{RC}}.$$

Since the MRSD expresses the average uncertainty relative to the reconstructed intensity values for the entire image it can be used as a quality metric for reconstructed images without the comparison to a ground truth image.

## Acknowledgements

The research leading to these results has received funding by the Deutsche Forschungsgemeinschaft (DFG), Research Unit FOR 5298 (iMAGO), subproject TP3 (GZ: PL 825/3-1). The development of MiROM has received funding from the Deutsche Forschungsgemeinschaft (DFG), Germany (Gottfried Wilhelm Leibniz Prize 2013; NT 3/10-1), as well as from the European Research Council (ERC) under the European Union's Horizon 2020 research and innovation program under grant agreement No 694968 (PREMSOT). We thank Dr. Serene Lee for her attentive reading and improvements of the manuscript.

## Author contributions

C.B. created and implemented the concept of BayROM and characterized the imaging framework. C.B. and M.K. designed and performed the experiments on the synthetic samples and adipose tissues. L.S.P. provided support on the implementation of the image reconstruction algorithm. V.N. provided support on optoacoustic microscopy. D.J. and M.A.P. supervised the whole study. All authors edited the manuscript.

## Conflicts of interests

V.N. and M.A.P. are founders and equity owners of sThesis GmbH. V.N. is a founder and equity owner of Maurus OY, iThera Medical GmbH, Spear UG and I3 Inc. The other authors declare no competing interests.

# 5 Extended data figures

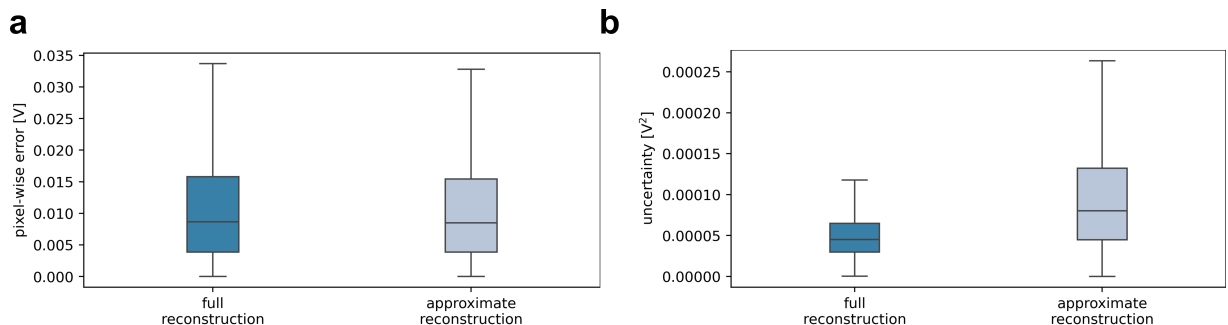

**Extended data fig. 1. Comparison between full and approximate reconstruction. a**, Comparison of pixel-wise error for full and approximate reconstruction. **b**, Comparison of pixel-wise uncertainty for full and approximate reconstruction. An approximate result can be obtained to speed up the reconstruction process by reducing computational complexity. Reconstruction based on 5 iterations of metric Gaussian variational inference (MGVI) and 16 samples drawn from the posterior distribution are considered full reconstructions, while approximate reconstructions were carried out using 3 iterations of MGVI and 8 samples drawn from the posterior distribution. The approximation (obtained in 2min 38s) has a similar error distribution as the full reconstruction (obtained in 10min 6s). However, the approximate reconstruction comes with less confidence (higher uncertainty) compared to the full reconstruction.